\documentclass[epj]{svjour}
\usepackage[latin1]{inputenc}
\usepackage{mathrsfs}
\usepackage{latexsym}
\usepackage{graphicx}
\usepackage{amssymb}
\usepackage{times}
\usepackage{psfrag}
\usepackage[dvips]{color}
\usepackage{amssymb}
\input colordvi
\def\thiog{$\mathrm{CdCr_{2-2x}In_{2x}S_4}$}
\def\thio{$\mathrm{CdCr_{1.7}In_{0.3}S_4}$}
\def\snif{$\mathrm{CsNiFeF_6}$}

\newcommand{\uu}[1]{\,\mathrm{#1}}

\begin{document}
\title{Off-equilibrium fluctuation-dissipation relation in a spin glass}
\subtitle{An experimental test for mean-field predictions}
\author{D. H\'erisson\inst{1}\thanks{\emph{Present address:} Department of Engineering Sciences--- Division of Solid State Physics, Uppsala University, 751 21 Uppsala, Sweden} \and M. Ocio\inst{1}\thanks{Deceased 21 December 2003}
}                     
%
%
\institute{Service de Physique de l'\'Etat condens\'e, CEA---centre de Saclay, Orme des Merisiers, 91 191 Gif-sur-Yvette cedex, France%
}
\date{Received: date / Revised version: date}
%
\abstract{We report new experimental results obtained on
the insulating spin glass \thio. Our experimental setup
allows a quantitative comparison between the
thermo-remanent magnetisation and the autocorrelation of
spontaneous fluctuations of magnetisation, yielding a
complete determination of the fluctuation-dissipation
relation. The dynamics can be studied both in the
quasi-equilibrium regime, where the fluctuation-dissipation
theorem holds, and in the deeply ageing regime. The limit of
separation of time-scales, as used in analytical
calculations, can be approached by use of a scaling
procedure.
\PACS{ {05.70.Ln}{Non-equilibrium and irreversible thermodynamics}
\and {75.50.Lk}{Spin glasses and other random magnets} \and
{07.20.Dt}{Thermometers} \and {07.55.Jg}{Magnetometers for
susceptibility, magnetic moment, and magnetisation measurements}
} 
} 
\maketitle
\section{Introduction}
\label{sec:introduction}

Despite their large diversity, glassy systems have  many dynamical properties in
common. In particular, a similar
ageing behaviour can be observed in polymers, gelatins, or
spin glasses \cite{Struik,Mydosh}. Stationarity cannot be
reached in these systems in experimental, or even in
geological times: they always remain out-of-equilibrium,
even when not submitted to any external perturbation.

During a long period, the theoretical activity was
concentrated on the study of the statics of glassy models.
With the nineties, began the time of theoretical dynamical
studies, first by numerical simulations, and then by
analytical results on specific mean-field models
\cite{Cugliandolo16,Cugliandolo01,Cugliandolo10}. From
these studies, new concepts appeared, generalising the well
known Fluctuation-Dissipation Theorem (FDT), which holds
for equilibrated systems \cite{Callen00,Kubo}.

In equilibrated systems with time translational invariance
(TTI), FDT can be used to measure the temperature in an
absolute way:
\begin{equation}
k_B T=\frac{\partial_{t_w}C(t-t_w)}{R(t-t_w)}
\label{eq:FDT}
\end{equation}
In this relation $C(t-t_w)$ is the  autocorrelation
function of an observable (for instance the magnetisation
$M(t)$) between two times, $t$ and $t_w$, and $R(t-t_w)$
the response function associated with a pulse of the
conjugate field at time $t_w$, $h(t)=\delta(t-t_w)$. These
quantities are two times quantities, but, as the system is
TTI, they depend only on the time difference $t-t_w$ .

Spin glasses never reach equilibrium, and the time
autocorrelation and the response function can not be
reduced to one-time quantities. Therefore, the temperature
cannot be defined on the basis of usual concepts.
Nevertheless, it has been shown that, in specific models
with low rate of entropy production, and using a
generalisation of the FDT relation, a quantity that behaves
like a temperature could be defined \cite{Cugliandolo10},
the ``effective temperature''. The effective temperature
for one given value of $C(t_w ,t)=C$ can be defined as:
\begin{equation}
k_B T_{eff}=\lim_{
\begin{array}{c}
{}_{t_w\rightarrow\infty}\\
{}_{C(t_w,t)=C}
\end{array}
}\frac{\partial_{t_w}C(t_w,t)}{R(t_w,t)} \label{eq:FDTg}
\end{equation}
The only difference between relations \ref{eq:FDT} and
\ref{eq:FDTg} concerns the domains of validity. The
generalised fluctuation-dissipation relation is valid for
stationary systems (simply, $C$ and $R$ depend only on $t-
t_w$ and $T_{eff}=T$), and it is also valid for every
systems in the limit of small rate of entropy production.
Glassy systems, in the limit of long waiting time are such
systems. Some experiments have been set up to measure this effective temperature using frequency measurements in glassy systems \cite{Grigera00,Bellon00,Bellon03}.

 To understand the meaning of the time limit in
equation \ref{eq:FDTg}, it is helpful to refer to the
so-called ``Weak Ergodicity Breaking'' (WEB) concept
~\cite{JPBx1}. WEB was introduced first in the study of the
dynamics of a random trap model very similar to the Random
Energy Model (REM)~\cite{Derridax0}. According to WEB
scenario, two different contributions can be identified in
the dynamics: a stationary one, corresponding to usual
equilibrium dynamics in a metastable state (and then not
relevant for ageing studies), and a second one, describing
the long term evolution between many metastable states,
which features the ageing properties.

This approach agrees well with an experimental fact: in
glassy systems, the relaxation function can be decomposed
in two distinct contributions \cite{Vincent00}:
\begin{itemize}
    \item The first one is independent of the age of the system
    (it depends only on the observation time $t-t_w$), and
    governs the dynamics for the shorter observation times. Many results
    in spin glasses showed that the most appropriate form for the decay is a power
    law with a small exponent, $\alpha\approx0.1$. This behaviour is
    consistent with the quasi-equilibrium noise power spectrum, which varies
    as $1/f^{1-\alpha}$. As this part is stationary, it should
    behave as in the equilibrated system:
    FDT should hold between the stationary part of the relaxation and the
    corresponding part of the autocorrelation, as shown in the section \ref{sec:raw}.

    \item The second one is non-stationary and decays approximately as a stretched
    exponential of the
    ratio $t/t_w$. This means that if $t_w \to 0$, this part tends to be instantaneous,
    and if $t_w \to \infty$, it becomes infinitely slow. This contribution can
    be rescaled using a re-parametrised (effective) time $\lambda(t)$. When plotted
    versus the effective time difference,
    all the non-stationary contributions measured with different waiting times
    merge very satisfactorily in one
    curve \cite{Vincent00}, showing that the same kind of dynamics persists along
    the whole experimental time range, as in Fig.~\ref{fig:scal}. Here, FDT cannot be of any help to link response and stationary parts.

\end{itemize}

The limit in the definition of the effective temperature  (equation \ref{eq:FDTg}) means that the two
contributions must be well separated, \textit{i.e.,} the
stationary dynamics must become negligible before the ageing
one starts to be effective. This situation is referred as
the ``time-scale separation limit", and the evolution of
any dynamic quantity should present a \textit{plateau} (in
log-scale of time) separating the stationary dynamics at
short times from the ageing one. Experimentally, this clear
separation of the two contributions is not observed.

In section \ref{sec:FDTthermometer}, a setup allowing the
measurement of magnetic fluctuations and the response to
the conjugate field will be described, and it will be shown
that it allows an absolute measurement of the temperature,
following Eq.~\ref{eq:FDT}. In section \ref{sec:FDR}, the
experimental procedure to study the ageing regime of a
spin-glass using this tool is described. The results allow
to check the validity of the effective temperature concept,
following Eq.~\ref{eq:FDTg}, and analysed according to
various models in section \ref{sec:discussion}.

\section{An FDT-based thermometer}
\label{sec:FDTthermometer}

In this section, a new experimental setup, designed to
measure quantitatively the relations between fluctuation
and response in magnetic systems will be described. It will
be shown that this setup works in fact as an absolute thermometer.\\

 Using FDT, expressed as in Eq.~\ref{eq:FDT} for instance, any
system allowing a quantitative comparison between thermal
spontaneous fluctuations of an observable and the response
to its conjugate field allows an absolute determination of
the temperature. The new experimental setup developed for
the studies of spin-glasses is first of all an absolute
thermometer, which should allow a determination of the
\textbf{thermodynamic} temperature of any equilibrated
magnetic system down to very low temperature. In a setup
completely dedicated to low temperature measurements, the
lowest temperature to be measurable should be below the
milliKelvin range.

\subsection{Noise measurements}
\label{sec:noise}

\begin{figure}
    \centering
    \includegraphics[width=0.75\columnwidth]{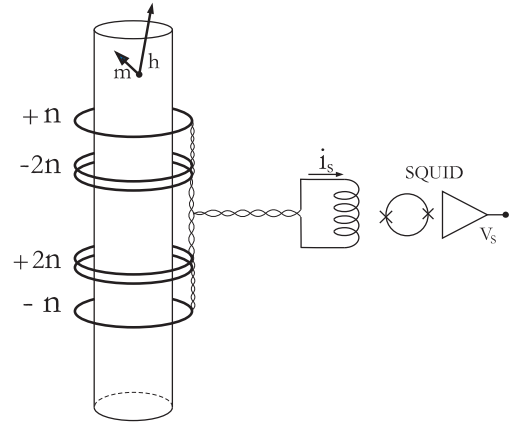}
    \caption{Schematic of the basic circuit for noise measurement. In order to maximise the coupling factor between the sample and the
Pick Up coil, a long cylinder ($4\uu{cm}$ long, $4\uu{mm}$ wide) is
used. The third order gradiometer being $2.2\uu{cm}$ long $5.5\uu{mm}$
wide, this size insures almost the best possible coupling factor, as
any contact between the PU-coil and the sample must be avoid to allow the
temperature regulation.}
    \label{fig:CircuitBruit}       
\end{figure}

The protocol of spontaneous magnetic fluctuations
measurements is quite simple: a thermalised sample is
introduced in a pick-up coil (PU), itself part of a
superconductive circuit involving the input coil of a
SQUID-based flux detector (see
Fig.~\ref{fig:CircuitBruit}).

Materially, the sample of cylindrical shape with diameter
and length  $5\uu{mm}$ and $40\uu{mm}$ respectively is
contained in a cylindrical vacuum jacket, part of a $^4$He
cryogenic equipment. The PU is wound on the jacket. The
sample itself is contained in a cylinder made of copper
coil-foil whose upper part is a copper sink with
thermometer resistor and heater resistor, thermally
connected to the $^4$He bath by a flexible copper link,
thus allowing temperature regulation above $4.2\uu{K}$. The
vacuum system involves a charcoal pump thermally connected
to the $^4$He through a thermal impedance. When cold, it
insures good thermal insulation; if heated, it allows to
inject $^4$He exchange gas thus thermally connecting the
whole sample to the helium bath temperature.

The difficulties of the measurement lie in the extreme
weakness of the signal of the fluctuations and the strong response
to external excitations: the typical amplitude of magnetic
fluctuations in our macroscopic \thio{ } sample corresponds
to the response to a magnetic field about $10^{-7}\uu{G}$.
Several magnetic shields ($\mu$-metal and superconductive)
are used in order to decrease the residual field at a level
of order $1\uu{mG}$, and to stabilise it. Furthermore, the
PU is built with a third order gradiometer geometry, which
strongly reduces the sensitivity to the time variations of
the ambient fields. In such conditions, and because of the
extreme sensibility of SQUID measurements, a satisfactory
signal/noise ratio can be easily obtained at short
time-scales, corresponding to correlation measurements with
time differences of few seconds. In order to study a
glassy system in the deep ageing regime, such time-scales
are not enough: one needs to measure the time
autocorrelation with time differences up to several 
thousand seconds. To suppress spurious drifts of the
measuring chain, further precautions are then needed:
stabilisation of the helium bath to avoid drifts of the
SQUIDS sensor, stabilisation of the room temperature to
avoid drifts in the ambient temperature electronics, etc.

It should be emphasised that the use of the third order
gradiometer in this experiment is quite different from the
most common use. Usually, gradiometers are used in
magnetometers where the sample is small compared to the
gradiometer size, and is placed in a non-symmetric
position; an homogeneous field is established, and the
unbalanced flux due to the magnetisation of the sample is
recorded.

\subsection{Response measurements}
\label{sec:response}

Already several years ago, a comparative study of the magnetic fluctuations and the conventional magnetic response was done using a setup similar to the one schematically described in figure~\ref{fig:CircuitBruit}
\cite{Refregier00}. This comparison could not be made
quantitative with a satisfactory accuracy: the coupling
factor of the sample to the detection system depends on the
PU geometry and is quite different in the noise setup and
in a classical magnetometer with homogeneous field. In
order to be able to compare \textit{quantitatively} the
results of both kinds of experiment, one has to eliminate
the effect of this geometrical factor in the comparison.
This can be done only if the coupling factor is the same in
both experiments. The way to achieve this can be illustrated very
simply. The fluctuations of the magnetisation are recorded
through a PU coil with a given geometry. According to the
reciprocity theorem, the flux fluctuations induced in the
coil are the fluctuations of the scalar product of local
magnetisation $\mathbf M \mathrm{dV} $ by the local field
$\mathbf{h}$ produced by a unit of current flowing in the
PU coil:
$$\Phi=\int_{sample} \mathbf{M}.\mathbf{h}\mathrm{dV} $$

\begin{figure}
    \centering
    \includegraphics[height=4cm]{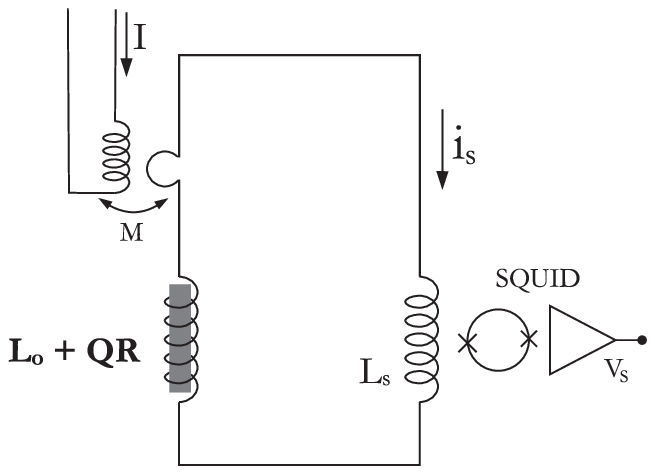}\\
    \includegraphics[height=4.5cm]{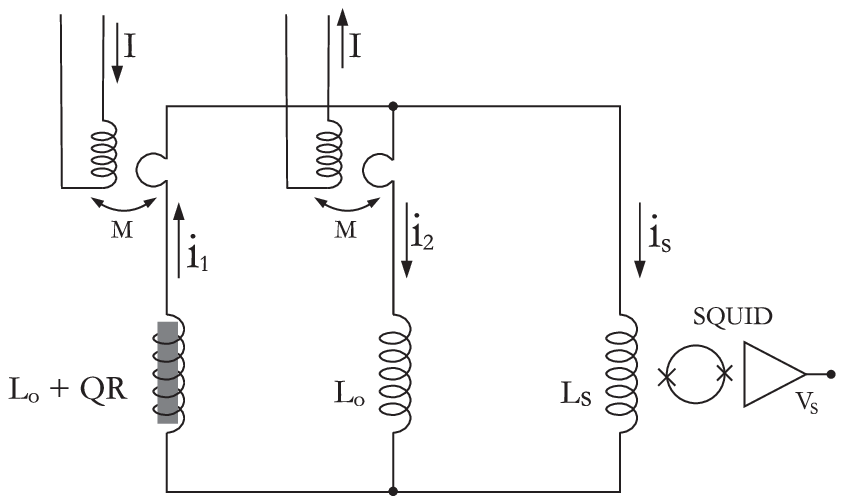}
    \caption{Schematic of the FDT circuits. \textsc{Top} : Basic FDT circuit. \textsc{Bottom} : The bridge configuration used.}
    \label{fig:Circuit}
\end{figure}

In our setup, the measured fluctuating observable is the
flux in the PU-coil. The conjugate quantity of this flux
is the current flowing through the coil, and thus, the
magnetic field conjugate of the sample magnetic moment is
the field produced by the PU-coil itself. If this field is
used as exciting field, then the fluctuation-dissipation
relation should remain the same for the macroscopic quantities
as for the microscopic ones. This is strongly different from
the situation where one tries to compare the results of
noise measurements to the results of classical response
measurements done in an homogeneous field: then the
coupling factor in both measurements has to be evaluated. A
way to use the PU-coil as field generator is the following.
A small coil coupled to an excitation winding with mutual
inductance $M$ is inserted in the basic superconductive
circuit (see Fig.~\ref{fig:Circuit}.a).

\subsection{Absolute thermometer}
\label{sec:thermometer}
\subsubsection{Fluctuation Dissipation relation}
\label{sec:absTh}
Here we will show that, for any given equilibrated system,
the validity of FDT on microscopic quantities results in
the validity of an ``effective FDT'' on measured
quantities. The factor $K$ which appears is setup
dependent, but sample independent.

When a magnetic sample is inserted into the PU coil, by the
reciprocity theorem, a moment $\mathbf{m}_i$ at position
$\mathbf{r}_i$ induces in the coil a flux
$\delta\Phi=\mathbf{m}_i\mathbf{h}_i$ . Therefore, the flux
in the coil due to the sample is given by
\begin{equation}
\Phi=\sum_i\sum_{\mu} m^{\mu}_i h^{\mu}_i \label{eq:phi},
\end{equation}
where $\mu$ indexes the spin components: $\mu=\{x,y,z\}$
for Heisenberg spins, $\mu=\{z\}$ for Ising ones, etc. We
suppose that the medium is homogeneous, the
components of the fluctuations are statistically independent and their spatial
correlations are much smaller than the scale of the PU:
\begin{equation}
\left<m^{\mu}_i(t')m^{\nu}_j(t)\right> =
\left<m(t')m(t)\right>\delta_{ij}\delta_{\mu\nu}.
\end{equation}
Then, the flux autocorrelation is given by
\begin{equation}
\left<\Phi(t')\Phi(t)\right>=\sum_{\mu}\sum_i {h^{\mu}_i}^2
\left<m(t')m(t)\right> = QC(t',t) \label{eq:QC}.
\end{equation}
The flux autocorrelation in the PU is thus the averaged one
site moment autocorrelation per degree of freedom
$C(t',t)$, multiplied by the coupling factor $Q$ determined
by the geometries of the PU field and of the sample.

On the other hand, the impulse response function of one
moment in the sample is given by
\begin{equation}
R^{\mu\nu}_{ij}(t',t)=\frac{\partial m^{\nu}_j}{\partial
h^{\mu}_i}=R(t',t)\delta_{ij}\delta_{\mu\nu},
\end{equation}
where $R(t',t)$ is the averaged one site response function
of the sample. If a current $i$ is flowing in the coil, the
flux on the coil due to the polarisation of the sample
reads
\begin{equation}
\Phi(t)=\sum_i\sum_{\mu} {h^{\mu}_i}^2 \int^t R(t',t) i(t')
dt'.
\end{equation}
Thus, the response function of the flux due to the sample
in the PU circuit is
\begin{equation}
R_{\Phi}(t',t)=QR(t',t). \label{eq:QR}
\end{equation}
The same coupling factor $Q=\sum_i\sum_{\mu} {h^{\mu}_i}^2$
determines the values of correlation and response of the
flux due to the sample.

Note that the term $h$ is the value of the internal field
in the sample, due to a unit of current flowing in the
coil. Therefore, $Q$ corresponds to the same demagnetising
field conditions in both measurements. Actually, $Q$ is
time dependent, since the internal field is $h=h_0\mu
(t',t)$ where $h_0$ is the field term generated by the coil
in vacuum and $\mu (t',t)$ is the time dependent sample
permeability, but the important point is that $Q(t',t)$ is
exactly the same in both experiments.

The above derivation is done in the context of a magnetic
system, showing that the measured quantities represent
those used in theoretical work, in which the single-site autocorrelation and response functions are computed, and averaged over the sample. Incidentally, an equivalent
derivation could be done for any system with magnetic
response, for instance the eddy currents in a conductor,
with the same result: the coupling factors are the same in
the fluctuations and the response measurements.

In the basic measurement circuit, Fig.~\ref{fig:Circuit}.a, the
total flux impulse response of the circuit to the current
$i(t')$ flowing in it is
\begin{equation}
R_L (t',t)=\sum L\delta(t-t')+Q(t',t)R(t',t), \label{eq:RL}
\end{equation}
where $\sum L$ is the total self inductance of the circuit.
Flux conservation in the (SC) PU circuit leads to
\begin{equation}
\Phi_{exc}(t)+\int_{-\infty}^t R_L(t',t)i(t')dt'=0,
\label{eq:conservphi}
\end{equation}
where $\Phi_{exc}(t)=MI(t)$ is obtained by injecting a
current $I(t)$ in the excitation winding. The conjugate
variable of the circuit current $i$ is the flux
$\Phi_{exc}$ injected by the excitation coil. In the case
of an ergodic sample, it is easy to show that, once FDT
applies to the fluctuations and response of the flux
induced in the PU, it applies also to the fluctuations and
response of the current flowing in the circuit. Thus,
\begin{equation}
\sigma_i(t-t')=\frac{1}{k_BT}C_i(t-t').
\end{equation}
The SQUID gain is $G=V_S/i$. Thus, if a current
$I(t)=I_0(1-\theta(t))$ is injected in the excitation coil,
the relaxation of the SQUID output voltage is related to the autocorrelation function of
its fluctuations by:
\begin{equation}
V_S(t)=\frac{1}{KT} \left<V_S(0)V_S(t)\right>.
\label{eq:SQFDT}
\end{equation}
where $K=\frac{G}{MI_0} k_B\,$. The system is an absolute
(primary) thermometer since, by measuring both the response
voltage to an excitation current step and the
autocorrelation of the voltage free fluctuations, it allows
a determination of the temperature whose precision (once a
sample with large signal is chosen) depends only on the
precision of the determination of the experimental
parameters $I_0$, $G$ and $M$.

The main drawback of the elementary measuring circuit
depicted above is that the response to an excitation step
involves the instantaneous response of the total self
inductance of the circuit (first term in the right hand
side ---R.H.S--- of Eq.~\ref{eq:RL}). In our case, both the
susceptibility of the sample and the coupling factor $Q$
are weak. The quantity to be measured,--- the second term
in the R.H.S of Eq.~\ref{eq:RL}---, represents a few percent
of the first one. Thus, a bridge configuration as depicted in
Fig.~\ref{fig:Circuit}.b has been adopted. Now, the main branch
involving the sample is balanced by an equivalent one
without sample. This second branch is excited oppositely,
in such a way that when the sample is extracted from the
PU, there is no response of the SQUID to an excitation
step. When the sample is placed into the PU, the response
of the SQUID is determined only by the response of the
sample. Nevertheless, now, the loop coupling factor of the
sample to the SQUID involves different self inductance
terms in both measurements, and one gets
\begin{equation}
K=\frac{G}{MI_0}\frac{L_0 +2L_S}{L_0}k_B
\end{equation}
where $L_0$ and $L_S$ are the self inductances of the PU
and of the SQUID input respectively, and the effect of the
sample has been neglected in the value of $L_0$. This adds
sources of error on the calibration since the self
inductance values are difficult to determine precisely.

\subsubsection{Calibration.}
\label{sec:calibrationT} The circuit as described above is a
thermometer, allowing the determination of the temperature,
$k_B T$. Nevertheless, it involves several home-made coils
whose self-inductances cannot be determined in their
experimental environment without large errors. This dramatically
limits the precision on the determination of
the temperature. A calibration was thus needed. For this,
the fluctuations and response of a high conductivity copper
sample were measured in the setup. This  high
purity ($99,999\%$) sample has a very low residual resistivity at low
temperature, obtained by annealing at high temperature
in oxygen atmosphere, thus reducing the density of magnetic
residual impurities. The sample has a cylinder shape,
$5\uu{mm}$ wide and $4\uu{cm}$ long. It was thermalised at
the temperature of the boiling $^4$He at normal pressure
($4.215\uu{K}$).

\begin{figure}[ht]
    \centering
    \psfrag{correlation1}[Bc][tc]{\scriptsize $C(\tau)$}
    \psfrag{correlation1}[Bc][tc]{\scriptsize $C(\tau)$}
    \psfrag{correlation}[cc][Bc]{\scriptsize Autocorrelation $C(\tau)$,$[\uu{V^2_{PP}}]$}
    \psfrag{relaxation}{\scriptsize $\sigma (\tau)$,$[\uu{V}]$}
    \psfrag{time}[cc][cc]{\scriptsize Observation time, $\tau$~$[\uu{s}]$}
    \
    \includegraphics[width=0.47\textwidth]{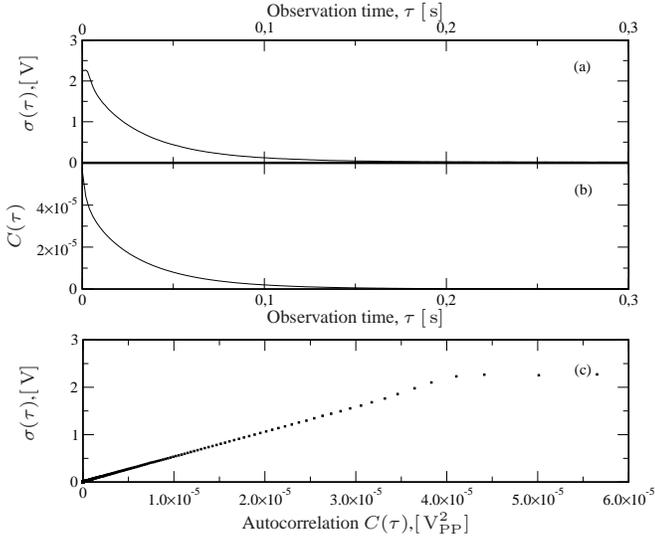}\\
    \caption{Measured relaxation (a)  and autocorrelation (b) function for the copper sample at $4.2\uu{K}$. (c) FD-plot, relaxation versus response, the observation time $\tau=t-t_w$ being used as parameter: the observed linear behaviour allows calibration of the system as thermometer, this slope being proportional to $1/T$, independently of the sample. The observed deviation from the linear behaviour at the shorter times is due to the effect of the low-pass filtering of the excitation--- which does not affect the correlation measurement.}
    \label{fig:cuivre}
\end{figure}

Since this equilibrated system is stationary, one can use
standard Fast Fourier Transform algorithms in order to
compute the autocorrelation function from a single record.
The  average of  the obtained autocorrelation function over
many successive records allows to reduce the noise level.

The relaxation function is obtained as the response to a
field step at $t_w$, and is only a function of $t-t_w$. As
the system does not have remanent magnetisation (the eddy
currents vanish in a finite time, a few tenth of a second), the
limit value of the response function is zero.

The measured relaxation and autocorrelation of SQUID
voltage are plotted in figure \ref{fig:cuivre}(a) and (b)
versus the observation time. The fluctuation-dissipation
diagram (FD-plot) is obtained by plotting the relaxation
versus autocorrelation, using the observation time as
parameter,  in figure \ref{fig:cuivre}(c) . The observed
linear behaviour is consistent with the FDT. As previously
shown, the slope between the relaxation and the response is
sample-independent, and proportional to $1/KT$. The
measurement on the copper sample, at a well known
temperature allows thus to determine the factor $K$. From
it, we can determine the temperature of any sample placed
in the gradiometer from the value of the slope of the
measured relaxation versus correlation curve. This can be
applied to any equilibrated system. For glassy systems, it
should allow an experimental determination of the effective
temperature.

\section{Fluctuation-dissipation relations in a spin-glass sample}
\label{sec:FDR}

The aim of this work is the study of the
fluctuation-dissipation relation in a spin glass. In this
section, we will emphasise the peculiarities of this
measurement, and then describe the procedure used to make
the analysis quantitative as well as the limits of this
procedure.

\subsection{Sample}
\label{sec:sample}
The knowledge acquired in previous magnetic noise
investigations on spin glasses was very helpful to choose a
good candidate for the present study. First, eddy currents
in metallic samples produce noise, as in the copper sample
used for calibration. This noise can be measured, but not
directly related to the spin dynamics. In order to avoid
this drawback, an insulating spin glass sample was chosen.

Measurements on \snif{ } have shown that this compound has
a stronger signal, and then a better signal to noise ratio
than any other insulating spin glass \cite{refregier04}.
Nevertheless it has a very strong ferromagnetic value of
the average interaction and its behaviour is far from the
``standard'' spin glass behaviour.

\thiog{ } was also extensively studied experimentally, by
classical susceptibility, magnetic noise, neutron
scattering \cite{Alba01,Albath,Alba02,Pouget02}. In this
series of compounds, the magnetic ions are
$\mathrm{Cr^{3+}}$, with low anisotropy.  The coupling is
ferromagnetic between first neighbours, and
anti-ferromagnetic between the second ones. The substitution
of $\mathrm{Cr^{3+}}$ by the non-magnetic
$\mathrm{In^{3+}}$ increases the relative importance of
anti-ferromagnetic coupling as compared to the
ferromagnetic one. The random dilution introduces disorder
and frustration, the basic ingredients leading to
spin-glasses. In the studies of the spin glass state,
\thio{ } is the preferred compound in this family. The
high concentration of $\mathrm{Cr^{3+}}$ allows this sample
to have a strong signal, but it is not high enough to reach the
percolation of the ferromagnetic order. With decreasing
temperature, finite sized ferromagnetic cluster formation
is observed. Close to $T_g$, these clusters are rigid, and
the interaction between them is random, with a weak
anti-ferromagnetic average. This clustering greatly
increases the signal, as the noise power of $N=N_0/n$
ferromagnetic clusters of $n$ spins is $\sqrt{n}$ times
stronger than the one of $N_0$ individuals spins.

\subsection{Experimental details}
\label{sec:details}

Glassy systems are not stationary; their dynamics depends
on two times, both referred to a crucial event, the
``birth'' of the system. In spin glasses, the birth time is
best defined by the time at which the final temperature is
reached, as soon as the end of the cooling procedure is
fast enough \cite{Jonason00}. A cooling procedure based
only on driving the sample holder sink temperature would
introduces non-negligible temperature gradients if the
cooling or heating rate is too high. This would lead to a
distribution of ages over the sample. In order to obtain a
more homogeneous temperature, the cooling procedure is as
follows:
\begin{itemize}
\item first, the temperature is slowly decreased from a reference
temperature $T_{ref}$ above $T_g$ to a temperature
$T_1\approx T_m+3\uu{K}$, where $T_m$ is the working
temperature;
\item then, by heating the charcoal pump, a small amount of He gas is
introduced, allowing a \textit{quick}
and \textit{homogeneous} cooling;
\item finally, the charcoal pump heating is switched off and the vacuum
surrounding the sample is restored, allowing the
temperature regulation at $T_m$.
\end{itemize}
\begin{figure}[htb]
  \centering
  \includegraphics[width=0.45\textwidth]{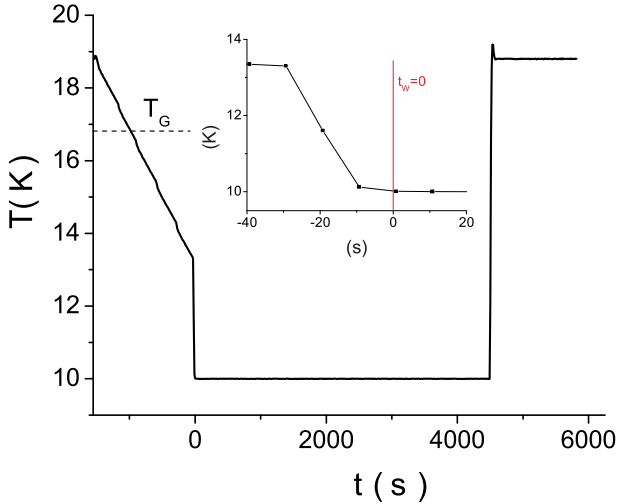}
  \caption{A typical thermal history of the sample for a $4,500\uu{s}$record at $10\uu{K}$. In inset, detail on the crucial part, the last $3\uu{K}$'s cooling.}
  \label{fig:descT}
\end{figure}
The first step (slow cooling, approx.
$0.25\uu{K.min^{-1}}$) does not introduce severe temperature
gradients, as the cooling is slow enough. This step cannot
be avoided, as the fast cooling by exchange gas can only be
used to decrease the sample temperature by few Kelvin 
without introducing too strong perturbations in the helium
bath. Previous studies on \thiog{ } show that the dynamics
is governed by the second, fast, cooling step, at least
on timescales shorter than $20,000\uu{s}$ \cite{Jonason00}.
This second step is obtained by heating the charcoal pump
during few seconds. As the gas surrounds the sample, the
resulting cooling is homogeneous. When the heating is
stopped, the charcoal absorbs the gas back, allowing the
regulation of temperature. The amount of gas used and the
duration of heating are adjusted in order to lower the
temperature exactly down to the working temperature. This
allows to cool the sample by $3\uu{K}$ in less than
$30\uu{s}$, keeping the temperature gradient negligible.
The birth time is taken as the instant when the sample temperature
reaches $T_m +15\uu{mK}$. This allows a precise, and
reproducible determination of it.

The measurement of the relaxation is straightforward. A DC
current is applied to the excitation coils at high
temperature, before the beginning of the quench procedure,
and switched off at $t=t_w$. The relaxation is then
recorded for $t>t_w$. The signal is measured before
applying the excitation: this determines the zero baseline
of the measurement. After relaxation, the sample is
re-heated to the start temperature in order to check the
stability of the baseline. Thus, in the measurement, both
the zero and field cooled (FC) levels are known.

Recording the fluctuations is even simpler, at least in
principle: no field is applied, the spontaneous
fluctuations of the signal are just recorded from the end
of the quench procedure and during a long enough time to be
able to compute all the desired $C(t_w,t)$. However, as the
system is not ergodic in the ageing regime, the
autocorrelation of the signal \textit{cannot} be evaluated
from a single record, as for an equilibrated sample. In
order to compute the autocorrelation function, an ensemble
average has to be done on successive equivalent records,
each one initiated by a quench. This does not only increase
dramatically the length of the experiment, but also the
difficulties of the acquisition, the ideal acquisition
conditions having to be kept during months instead of
hours. This complication has nevertheless an advantage: by
averaging, it allows to make the separation between a
systematic spurious signal and the signal of the fluctuations. In
our results, the systematic signal is of the same order of
magnitude as the fluctuations themselves. It corresponds
to the drift of the SQUID due to the continuous decrease of
the He-level, and to the global response of the sample to
the residual field during the cooling procedure. The
average of the signal over records gives the zero, and the
sample fluctuations signal is then given by:
\begin{equation}
m(t)=M(t)-\left< M(t)\right>
\label{eq:signal}
\end{equation}
The autocorrelation is then evaluated from its definition:
\begin{equation}
C_1(t,t_w)=\left< m(t_w)m(t)\right>
\label{eq:Cexp}
\end{equation}
In order to obtain a small statistical error, a very huge
number of records is needed. As each record length is about
few hours, the number of records is limited to about 300,
and the average over the records is not enough to obtain a
satisfactory ratio between the statistical error and the
signal. As the autocorrelation function should evolves
smoothly for both variables, $t_w$ and $t$, the
autocorrelation function computed following eq.~\ref{eq:Cexp} is averaged over small time
intervals of both variables :
\begin{equation}
C_{avg}(t_0,t_{w_0})= \overline{C_1(t,t_w)} \vert _{t_w\in
[t_{w_0}\pm \epsilon_{t_{w_0}}],\,t\in [t_{0}\pm
\epsilon_{t_0}]} \label{eq:MoyTempC}
\end{equation}
with
\begin{equation}
\epsilon_{t}=0.05t\ll t.
\label{eq:epsilon}
\end{equation}
The criterion used (Eq.~\ref{eq:epsilon}) is a compromise between the need of statistics in order to obtain a low enough statistical noise, and the requirement to keep $\epsilon$ as small as possible to be able to capture as precisely as possible the non-equilibrium dynamics.

\subsection{Correlation offset.}
\label{sec:calibrationII}

In principle, our experimental procedure, involving many
realisations of the same experiment, allows to compute the
autocorrelation function of the magnetisation following its
exact definition, and thus exactly. Nevertheless, in
reality, this would be the case only if external sources of
noise were negligible, not only in the correlation
time-scale under study, but also in the time-scale of one
complete record. This means that the external noise should
be controlled not down to $1\uu{mHz}$ as in our experiment,
but at least down to frequencies as low as few
$0.01\uu{mHz}$, which is quite impossible. The result is
that the computed correlation $C(t_w ,t-t_w)$ contains an
offset practically independent on $t-t_w$ but randomly
dependent on $t_w$.

As the setup is a calibrated thermometer, the temperature
can be extracted from the derivative of the $\chi(C)$
curves in experimental units. In order to obtain the
FD-plot, this is not enough. For a normalisation of the
data, one needs to know the zero reference level of the
response and of the correlation. In the case of the copper
sample data, where the eddy currents producing the signal
have a finite and experimentally accessible lifetime, this
calibration is trivial: relaxation and autocorrelation
functions decrease to zero after a few seconds. In the spin
glass case, normalisation of the relaxation is simple since
the zero level and the FC level are determined during the
measurement. For the correlation, things are not so easy.

As seen above, at a given temperature, the
correlation curves for different $t_w$ are shifted between
each other by a random unknown offset. Nevertheless, it is
possible to normalise the data by taking as the origin of
each curve the square of the measured value of the first
point, $C_{avg}(t_w,t_w)$. Due to the elementary
measurement time constant, this term corresponds to an
average over $t-t'$ about $10^{-2}\uu{s}$, \textit{i.e.}, a range of
$(t-t')/t'$ corresponding to the stationary regime where
all curves must merge. Thus, the following quantity is
computed:
$$C(t,t_w)-C_0=C_{avg}(t,t_w)-C_{avg}(t_w,t_w)$$
The ``individual" offset is now replaced by a ``global''
one, $C_0$, which should apply simultaneously to any
measurement done at a given temperature.

Then, the best way to normalise our data could be to extract
$C(t,t)$ from some other measurement, and to be able to
convert it in the ``experimental units''. Neutron
diffraction experiments are now under way in order to
measure this quantity. $C(t,t)$ can also be extrapolated
from high temperature measurements (above $T_g$) to low
temperatures (below $T_g$), or deduced from some other
quantities. Then a complete ---but model-dependent---
determination of the autocorrelation can be obtained,
allowing to obtain the FD-plot. Anyway, even if the
hypothesis used to obtain this complete determination of the
autocorrelation were not realistic, some characteristics of
the FD-plot would not be affected. The temperatures,
effective or not, are measured from the slope between
relaxation and correlation in the experimental units. They
will not be modified by the normalisation procedure, whose effect is just
 to suppress an offset.

\section{Discussion}
\label{sec:discussion}

In this section, the results of the measurements done at
several temperatures in \thio{ } will be analysed
following the line of the method described above.

\subsection{Raw measurement}
\label{sec:raw}

\begin{figure*}
    \begin{tabular}[c]{c c}
        \includegraphics[width=0.45\textwidth]{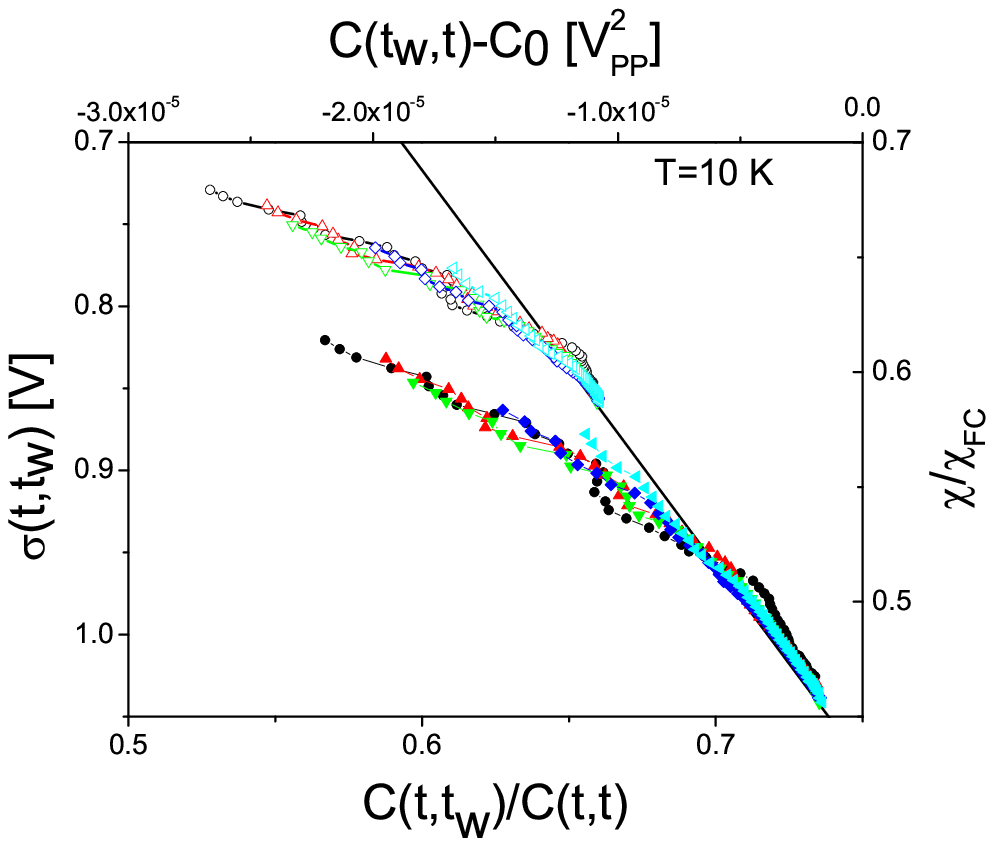} &
        \includegraphics[width=0.45\textwidth]{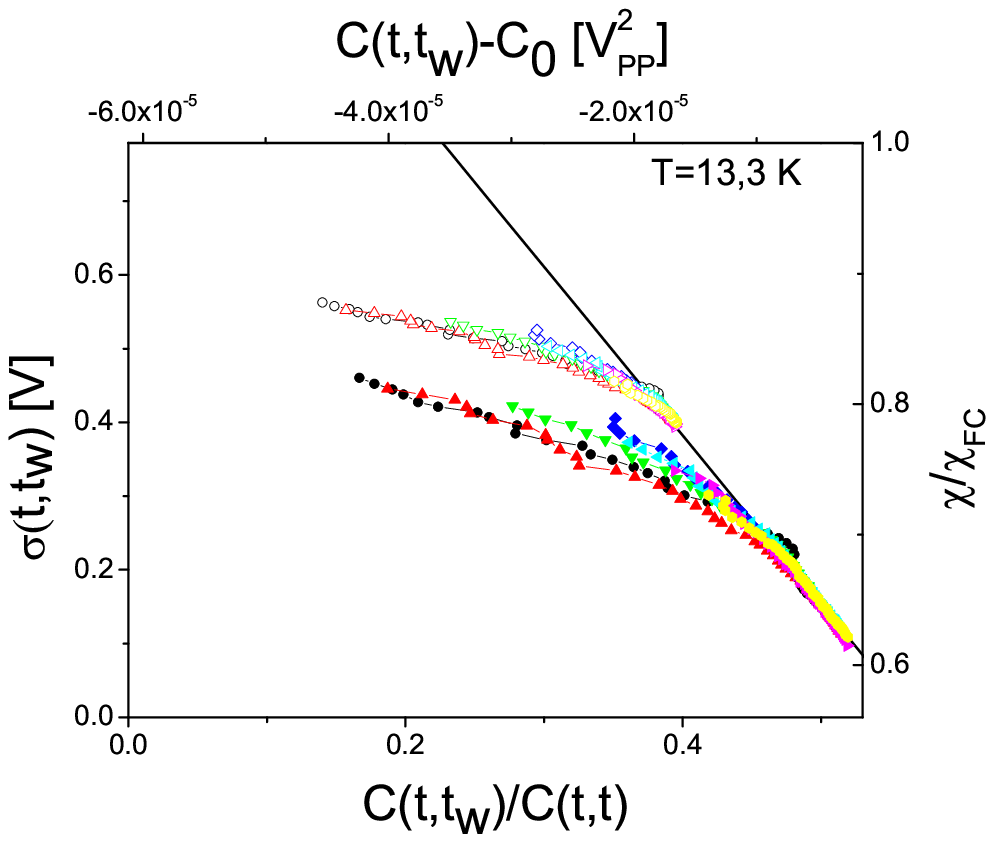}\\
        \includegraphics[width=0.45\textwidth]{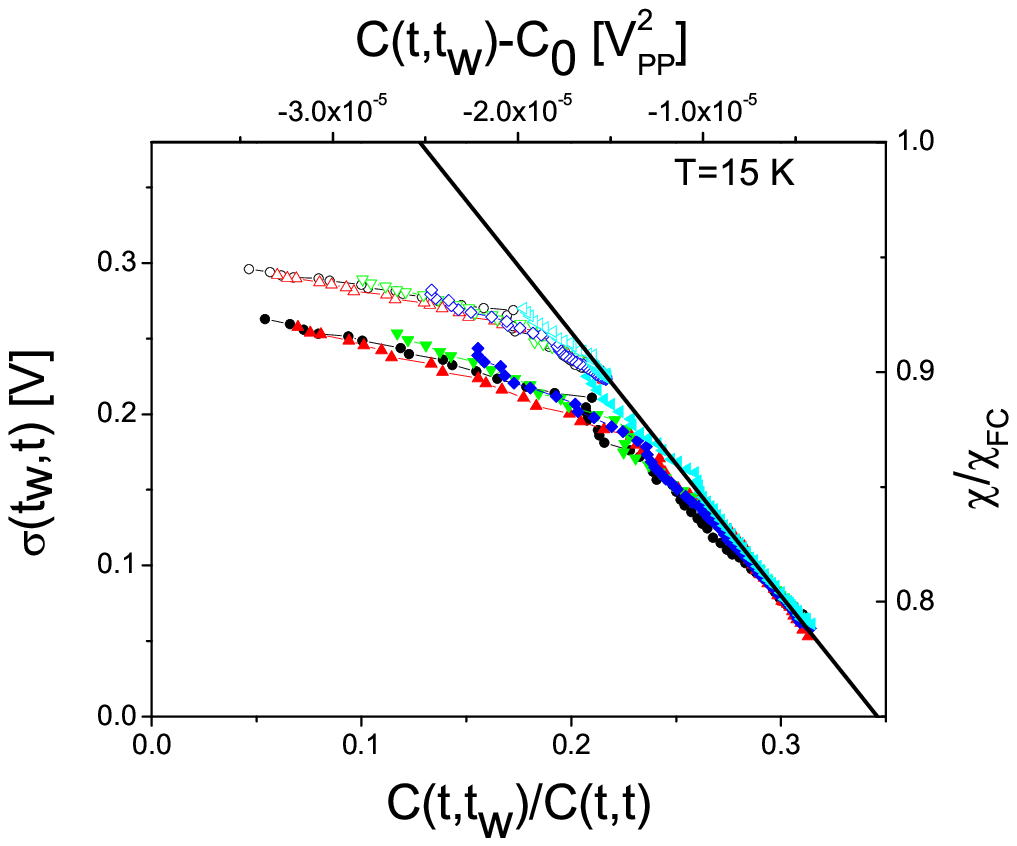} &
        \includegraphics[width=0.45\textwidth]{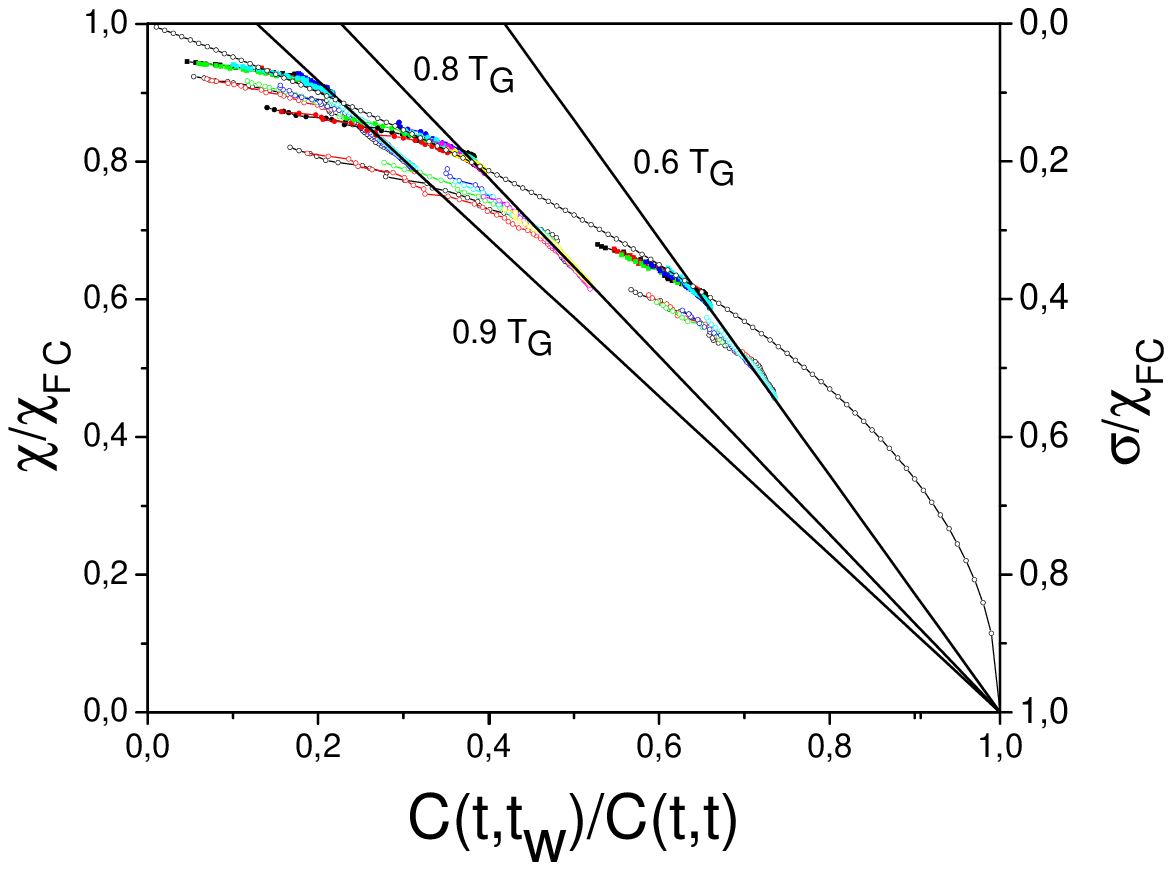}\\
    \end{tabular}
    \caption{Raw results (full symbols) and ageing part (open symbols)
    deduced from the scaling analysis for the three investigated
    temperature, $T=10, 13.3, 15 \uu{K}$. The straight lines have
    the FDT-slope obtained from the copper calibration for each
    temperature of the formalisation bath, and start all from
    the point $(\sigma/\chi_{FC}=1,C/C(t,t)=1)$. The correlation
    offset is adjusted following the ``PaT hypothesis''
    (cf section~\ref{sec:Continuous}) The different curves span
    the waiting times studied: $\circ :t_w=100\uu{s},\,\textcolor{red}
    {\triangle} :t_w=200\uu{s},\,\textcolor{green}{ \triangledown}
    :t_w=500\uu{s},\, \textcolor{blue}{\diamond} :t_w=1000\uu{s},\,
    \textcolor{cyan}{\triangleleft} :t_w=2000\uu{s} $ and, only for
    $T=13.3\uu{K}$, $\textcolor{magenta}{\triangleright} :t_w=5000\uu{s},\,
    \textcolor{yellow}{ \circ} : t_w=10000\uu{s}$.The last plot represent
    the complete FD-plot for each previous measurement measurements
    reported on the same graph, the smooth curve corresponds to equation \ref{eq:master_extended}, with an exponent $B=0.47$. }

    \label{fig:CuKu}
\end{figure*}

Figure \ref{fig:CuKu} displays the values of $\chi(t_w
,t)/\chi_{FC}= 1-\sigma(t_w ,t)/\chi_{FC}$ plotted versus
$C(t_w ,t)$ for several values of $t_w$ and using $t-t_w$
as parameter. The three graphs correspond to the three
temperatures 10, 13.3 and $15\uu{K}$. A first observation
is that a linear regime exists between relaxation and
correlation for all the temperatures and waiting-time
investigated. This regime corresponds to the shorter
observation times. In the figures, the straight lines
represent the FDT slope as calculated from the values of
calibration factor $K$ and of the temperature: in this
regime, the relation between relaxation and correlation
follows the FDT. Thus, this regime can be extrapolated from
the shorter experimental observation-time down to the
microscopic time-scale.  This extrapolation at short time
should reach the starting point of the FD-plot:
$C(t_w,t_w)$ corresponds to $\chi(t_w,t_w)=0$. As no long
term memory is observed in spin-glasses, $C(t_w,+\infty)=0$
should also correspond to $\chi(t_w,+\infty)=\chi_{FC}$,
but the extrapolation to this point is not obvious at all,
as the (unknown) ageing regime should be extrapolated.

\subsection{Scaling procedure}
\label{sec:scal}

The raw results show a waiting time dependence which can be
easily explained. The main theoretical predictions
correspond to the approach of the limit $t_w\rightarrow
\infty,\,C(t_w,t)=C$ (WEB). In this case, the
stationary and the ageing parts of the dynamics evolve on
distinct time-scales, yielding a separation of both
dynamics.
Experimentally, such a separation is not accessible since
the waiting times are finite. In order to separate both part of the dynamics, the entanglement between both part should be described.
The simplest way to combine these two contributions is to
add them. For instance, if one considers the relaxation
$\sigma$ observed after a unitary field step at time $t_w$:

\begin{eqnarray}
  \nonumber  \sigma(t_w,t)=(1-\Delta).\sigma_{stat}(t-t_w)+\\ \Delta.\sigma_{ageing}\left(\lambda(t)-\lambda(t_w)\right)
    \label{eq:Scal+}
\end{eqnarray}

In this equation, all the different $\sigma$ are normalised
to unity. This relation is obviously valid in the limit of
separation of time-scales, and is the most commonly used in
theoretical approaches, but it is counter-intuitive as
shown by the following two thought-experiments:

\begin{figure}
	\centering
	\includegraphics[width=0.45\textwidth]{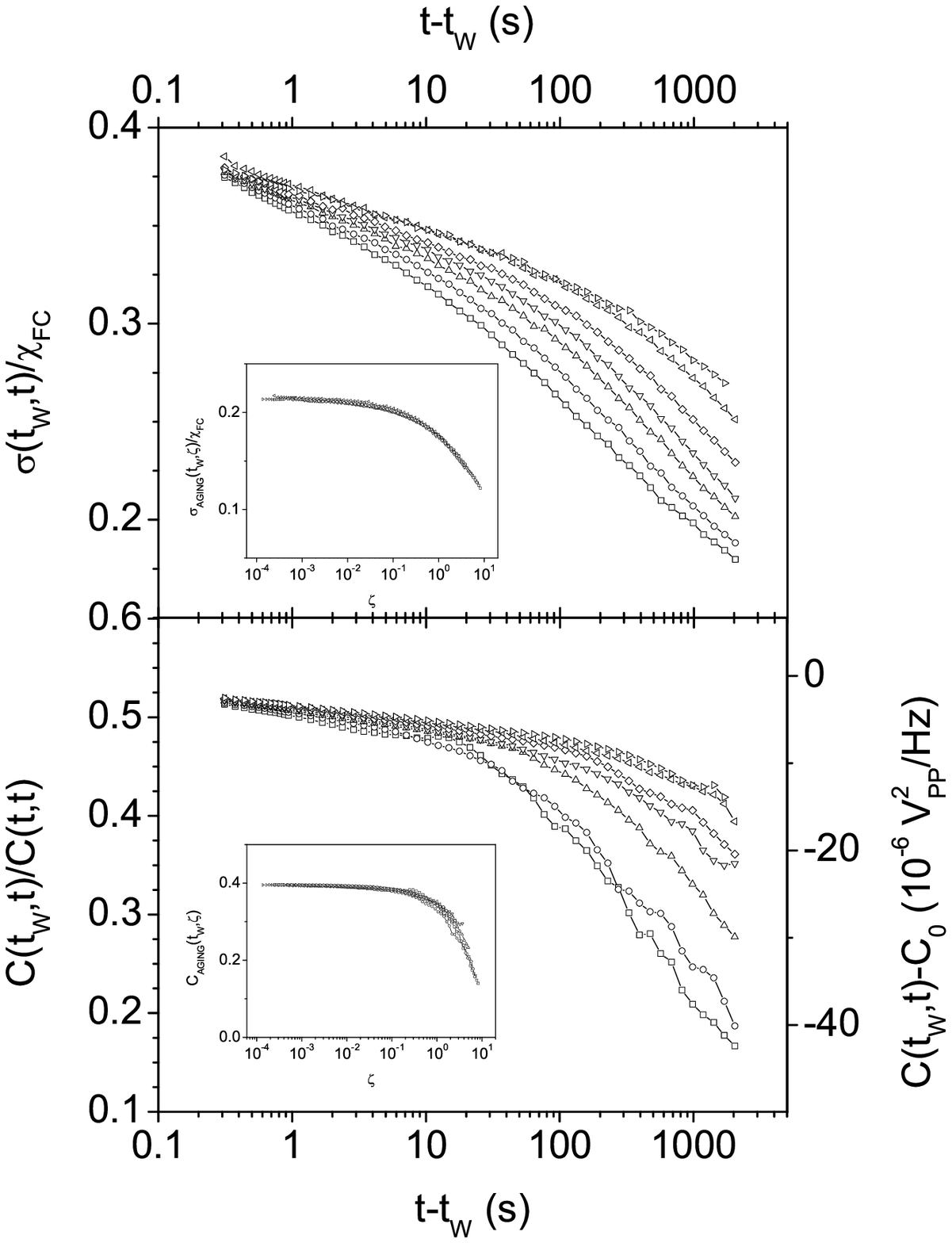}
	\caption{Relaxation (\textsc{Top}) and autocorrelation (\textsc{Bottom}) functions  recorded at $13.3\uu{K}$. The
	different curves correspond, from bottom to top, to $t_w=100$, 200, 500, 1 000, 2 000, 5 000 and $10\,000\uu{s}$.
	In insets, the respective ageing parts, deduced by the scaling analysis developed in section \ref{sec:scal}}
	\label{fig:scal}
\end{figure}

\begin{itemize}
    \item In the first one, a glassy system is quenched at a temperature
    below $T_g$ at time $t=0$, and at $t=0^+$, a field is applied. This experiment is
    usually though as being equivalent to the ``Field-cooled''
    procedure, in which the field is applied \textit{before} the quench.
    Experimentally, the Field Cooled magnetisation is strikingly stable. In
    the additive formulation, the predicted
    behaviour is the following: first, an instantaneous variation due to
    the ageing part, and then a slow variation up to the FC value,
    as the system approaches equilibrium. Thus, the field-cooled
    magnetisation should not be as stable in time, as what is
    observed experimentally.

    \item The second problem arises when thinking about some finite $t_w$
    experiments, but with (very) huge time differences,
    $t\rightarrow\infty$. Ageing and stationary parts are usually described
    as stretched exponential (with characteristic time of order $t_w$) and
    power-law with small exponent respectively.
    For finite $t_w$, after a finite time, the only remaining contribution
    to the dynamics would come from the stationary part, and FDT would be recovered.
\end{itemize}

If the time-scales are not well separated, it seems
intuitively that the two different contributions must be
more entangled than the result of a simple addition.
Another (more realistic) way to mix the two parts together
is to consider a ``multiplicative'' combination, which can
be written as:

\begin{eqnarray}
   \nonumber \sigma(t_w,t)=\left[(1-\Delta).\sigma_{stat} (t-t_w)+\Delta \right]\times\\
    \sigma_{ageing}\left(\lambda(t)-\lambda(t_w)\right)
    \label{eq:ScalX}
\end{eqnarray}

In the limit of time-scales separation, equations
\ref{eq:Scal+} and \ref{eq:ScalX} are equivalent. Moreover,
it is easy to show that the problems raised in the two
cases discussed above disappear. Thus, as this formulation
is at least as justified as the ``additive'' one, and seems
less counter-intuitive to the authors, it will be preferred
in the following.

Experimentally, separation of timescales is not accessible since
the waiting times are finite. In order to separate both part of the dynamics, a scaling analysis, as described by equation \ref{eq:ScalX} and illustrated by Fig.~\ref{fig:scal} is applied on both relaxation and correlation measurements,
 within the following constraints:
\begin{enumerate}
    \item  \label{it:Calpha}In the relaxation, the stationary
    part is described by a power-law decay; its exponent $\alpha$ is
    extracted from the decay of the noise power-spectra measured
    on the same sample, at the same temperature, in the quasi-equilibrium
    regime obtained after a very long waiting at the working temperature
    (typically 15 days) \cite{Refregier03}.
    \item \label{it:Cteff} In the non-stationary regime, the effective time is given by
    $\lambda=\frac{t^{1-\mu}}{1-\mu}$. The value of the sub-ageing exponent, \textit{i.e.},
    $\mu$ is in the range $0.85-0.9$ \cite{Alba01}. $\mu$ and the relative
    amplitude of the  ageing part,
    \textit{i.e.}, $\Delta$,  are chosen to obtain the best
    rescaling
    of the
    relaxation curves once the stationary part has been subtracted.
    \item \label{it:Ccorr}The stationary part of the autocorrelation
    function is then evaluated using FDT and the stationary part of the
    relaxation. The ageing part of the autocorrelation can then be deduced
    from equation \ref{eq:Scal+} or \ref{eq:ScalX}.
\end{enumerate}

The resulting FD-plots are displayed in open symbols in the
diagrams of figure~\ref{fig:CuKu} for the ageing part (by construction, the stationary part follows the FDT line). For all the investigated
temperatures, the ageing part starts with a slope very
close to the FDT one. This may reflect the imperfections of
our decomposition between the stationary and the ageing
parts. Using the additive scaling, this effect is even more
pronounced.

The multiplicative scaling has a main drawback: it
can not be used without the knowledge of the amplitude of
the concerned physical quantity. As previously discussed in
section~\ref{sec:calibrationII}, for the correlation this
value must be determined indirectly, and is model
dependent; thus the scaling is also model-dependent. The
additive form of the scaling can be applied without any
amplitude parameter. However, the dependence of the scaling
on this parameter is weak, the results obtained from
different models are indistinguishable from each other.

The ageing part of the correlation function is found to
follow remarkably the scaling-law used for the ageing part
of the response, resulting in a very weak systematic $t_w$-dependence of the FD-plots.
Thus, the equation \ref{eq:Scal+} (or
\ref{eq:ScalX}) can be written for the correlation, with the amplitude parameter
$\Delta$ replaced by $q_{EA}$, the usual Edwards-Anderson
order parameter \cite{EA}, which is defined as the remaining part of the autocorrelation for an equilibrated spin glass after an infinite waiting time. As a consequence, the FD-plots
are determined by a single parameter, the effective time
difference: the limit FD-plots, corresponding to the ideal
separation of the stationary and ageing regime is
independent of the age of the system, as supposed in
theoretical works.

\subsection{Comparisons with some models predictions}
\label{sec:Comparisons}

Depending on the models, some remarkable features of the
FD-diagrams are predicted. The analysis of the FD-diagrams
may help to check the validity of the models used to
interpret the glassy behaviour found in \thio.

\subsubsection{Domain growth}
\label{sec:Domain}

In domain growth models, as in any replica symmetric
models, the FD-plot should be quite simple in the limit of
time-scale separation. For an infinite waiting  time $t_w$, the quasi-equilibrium relaxation should go down to zero. A FDT-behaviour should then describe all
the response, governed by the
single-domain response and the scaling approach used in this paper should give $(q_{EA}\neq 0,\Delta = 0)$.  Thus, the remaining part of the
diagram, an horizontal line, should correspond to an
infinite effective temperature
\cite{Barrat02,Corberi00}. This description does
not coincide with the FD-diagram shown in figure
\ref{fig:CuKu}, even after
the separation of time-scales obtained by scaling.

Anyway, a more refined approach as in \cite{Yoshino00} is not excluded, in which dynamics is described introducing  a crossover region in between the quasi-equilibrium region ($C > q_{EA}$) and a purely ageing region characterised  ``dynamical order parameter'' $q_D$ for $C < q_D < q_{EA}$. This approach could explain the ``early'' departure from the FDT regime. This departure should be $t_w$-dependent, but this dependence may be hidden by a too small range of waiting times explored (as well as a too weak exploration of the ageing regime).

\subsubsection{1-step replica symmetry breaking}
\label{sec:1step}

In \thio , one of the best realisation of an Heisenberg
spin-glass, it has been shown that the scenario of the
chiral spin glass could be relevant \cite{Petit00,Petit01}.
Such model belongs to the 1-step replica symmetry breaking
(1-RSB) models family
\cite{Kawamura02,Kawamura03,Hukushima00}.

In 1-RSB case, the ageing regime is described by a unique
effective temperature, finite and  strictly greater than
the thermalisation temperature. By considering a normalised
FD-plot, it is easy to show that the value of $q_{EA}$ can
be deduced from the value of $T$, $T_{eff}$ and $\gamma={1- \Delta \over \Delta}$, the ratio
between the  stationary and the ageing  part of the
relaxation,  which is
experimentally accessible:
\begin{equation}
    q_{EA}=\frac{1}{1+\gamma.{T\over T_{eff}}}
    \label{eq:qea}
\end{equation}

\begin{table}
  \centering
    \begin{tabular}{llll}
        \hline\noalign{\smallskip}
        $T\,[\uu{K}]$ & $T_{eff}\,[\uu{K}]$ &$q_{EA}^{1-step}$&$q_{EA}^{PaT}$  \\
        \noalign{\smallskip}\hline\noalign{\smallskip}
        10 & $28\pm6$  & $0.65 \pm 0.05$& 0.63\\
        13.3 &$50\pm 10$&$ 0.45\pm0.05 $& 0.37\\
        15 & $80\pm20$ & $ 0.36\pm0.07$ & 0.21\\
        \noalign{\smallskip}\hline
    \end{tabular}
    \caption{Values of $T_{eff}$  as obtained in an 1-step replica symmetry breaking scenario and the corresponding Edwards-Anderson order parameter $q_{EA}^{1-step}$ (cf section \ref{sec:1step}). The values of $q_{EA}^{PaT}$ are deduced from the PaT \textit{ansatz} (cf. section \ref{sec:Continuous}).}
  \label{tab:qEA}
\end{table}

As the experimental setup is a calibrated thermometer, it
allows an absolute determination of the temperatures. The
determination of $T$ and $T_{eff}$ extracted from the slope
of the stationary and the ageing part respectively allows
the complete determination of the offset $C_o$. The
obtained values of $T_{eff}$ and $q_{EA}$ are reported in
table~\ref{tab:qEA}. The separation by scaling between
stationary and ageing part being far from perfect, the
uncertainty on the determination of $T_{eff}$, and
consequently on $q_{EA}$ is quite large. The choice of
a scaling procedure also influences  the results (the
previously reported value for $T_{eff}\approx 30 \uu{K}$
for a thermalisation temperature of $13.3\uu{K}$ was
obtained by an additive scaling analysis of the data
\cite{herisson00}). The results of table \ref{tab:qEA} can
be compared with the results of simulations done on a
weakly anisotropic spin glass model by Kawamura
\cite{Kawamura09}. In this work, it was found that $\chi$
depends linearly on $C$ in the ageing regime. The effective
temperature was found to be independent of the temperature
of the thermalisation bath. This independence does not
appear in our data, but, maybe, it can be due to the
extremely low anisotropy used in the simulations. \thio~is
known to be an Heisenberg spin glass with a non negligible
anisotropy, which as been found to be five times stronger
than in the canonical $\mathrm{Ag\underline{Mn}}$ spin
glass.

\subsubsection{Continuous replica symmetry breaking}
\label{sec:Continuous}

In continuous replica symmetry breaking ($\infty$-RSB)
models, as the Sherrington-Kirkpatrick (SK) model
\cite{SK,SKb}, the Parisi order parameter is a continuous
function between 0 and $q_{EA}$ \cite{SGTAB,replica3}.
Links between statics and dynamics imply that the
corresponding effective temperature is a smooth and not
trivial function of the autocorrelation :
\begin{equation}
    \frac{1}{T_{eff}}(\mathscr{C})=\frac{1}{T}\int_0^\mathscr{C}P(q)\mathrm{d}q
    \label{eq:Teff}
\end{equation}

Then there is no trivial relationship between $q_{EA}$ and
the measured quantities. It is nevertheless possible to to
progress further if the studied compound is a canonical
spin-glass. In these systems, the FC susceptibility is
purely paramagnetic at high temperature, following equation
\ref{eq:SGHT}, and below $T_g$, its value is temperature
independent:
\begin{eqnarray}
    \chi_{FC}(T>T_g)=C(t,t)/k_B T
    \label{eq:SGHT}\\
    \chi_{FC}(T<T_g)=C(t,t)/k_B T_g
    \label{eq:SGBT}
\end{eqnarray}

The lower the concentration of magnetic ions in the
canonical sample, the smaller the probability of spins
clustering and the better the validity of the above
relations \cite{nagata00}. Thus, the value of $C(t,t)$ can
be straightforwardly derived from susceptibility data. The
canonical behaviour is also observed or imposed in
theoretical models, and known as the Parisi-Toulouse
Hypothesis \cite{PaT}. This ``PaT'' hypothesis implies that
the FC response is temperature independent, as observed in
diluted spin glasses.

In samples with high concentration in magnetic sites,
deviations from the simple canonical behaviour are observed,
as well as the formation of clusters of spins. At low
temperatures,
the response of single spins is no more observed, but the
response of some rigidly coupled groups of spins. For
macroscopic quantities, this is equivalent to the response
of fewer renormalised spins. In \thio, which has a mean
coupling constant strongly ferromagnetic
($\Theta=100\uu{K}$ \cite{Albath,Pouget00}), this
clustering may explain that at low temperature, but above
$T_g$, the compound behaves as a compound with
antiferromagnetic average of couplings. A standard
Curie-Weiss law description around $30\uu{K}$ gives a mean
coupling characterised by $\Theta\approx-9\uu{K}$ \cite{alba00}.
Such a description, with a non-trivial $\Theta(T)$
variation, should be associated with a non-trivial, but still
smooth, function $C(t,t;T)$. As informations on the
variations of $C(t,t;T)$ are lacking we propose to consider
that relations \ref{eq:SGHT} and \ref{eq:SGBT} are still
valid in the general case. This is a strong hypothesis
since it amounts to consider that the temperature variation
of $\chi_{FC}$ is due only to the temperature variation of
${C(t,t;T)}$. One can write:

\begin{eqnarray}
   \frac{C(t,t;T)}{k_B T} = \chi_{FC}(T<T_g) \frac{T_g^*}{T}
   \label{eq:HPPaT}
\end{eqnarray}

A smooth behaviour of $C(t,t;T)/k_B T$ around $T_g$ can be
obtained using $T_g^*=17.2\uu{K}$ in the formula .

This \textit{ansatz} gives an  access to the unknown offset
of the autocorrelation: i) the starting point of the
FD-plot, corresponding to
$[C(t_w,t_w;T);\,\sigma(t_w,t_w)]$ is completely defined,
ii) $C(t_w,\infty;T)$ corresponds to the point where the
FDT line reaches the level given by $\chi_{FC}(T<T_g)
\frac{T_g^*}{T}$. Then the FD-graph can be plotted in
reduced units as displayed in figure~\ref{fig:CuKu}d. In
this plot, the starting point ($C=1$) and the end point
($C=0$) are temperature independent. Furthermore, it has
been shown (for some mean field models, and approximately for the SK model) that not only these points but also all the ageing
part of the plot is temperature independent \cite{Cugliandolo17}. It is
conjectured that this can be still valid in short range
models \cite{Franz02,Marinari00,Marinari02}. This property is particularly interesting: it
allows, by measurements at several temperatures, to obtain
the whole ``master'' curve describing the ageing behaviour,
even if each set of data spans a limited portion of the
correlation. This feature has been already used to obtain
the master curve from response data, assuming that the
separation of time-scales is reached in usual
susceptibility measurements \cite{Cugliandolo00,Zotev01}.

In the SK model at small $C$, it can be shown that
the master curve should behave as \cite{Marinari00}:

\begin{equation}
  \chi(C)=\sqrt{1-C}
  \label{eq:master}
\end{equation}

For correlations close to zero, the slope of the FD-plot,
$X(C)=\int_0^C{P(q)}\mathrm{d}q$, should asymptotically
reach zero, as $P(0)$ is known to have a finite value in
the continuous RSB case.

Equation \ref{eq:master} can be  generalised by allowing
any exponent different from $0.5$:

\begin{equation}
  \chi(C)=(1-C)^B
  \label{eq:master_extended}
\end{equation}

Such a curve can be easily superimposed to our experimental
results. Using a coefficient $B=0.47$, a single curve can
describe the ageing regime at all the investigated
temperatures and for the data close to $q_{EA}(T)$. Both
the value of $q_{EA}(T)$ and the effective temperature
close to $q_{EA}$ seem to be well described by
Eq.~\ref{eq:master_extended}.

For $C \ll q_{EA}$, at each temperature, the experimental
points deviate from relation \ref{eq:master_extended}. This
cannot be due only to the smaller signal to noise ratio at
the longest timescales, since the effect seems to be more
pronounced at the highest temperature, where the sample
signal is the strongest.

A possible explanation is that the scenario with continuous
replica symmetry breaking should be  associated with a
continuous distribution of timescales describing the
system. As the limit of separation of timescales is not
reached in our results, the ageing regime itself is a
combination of many timescales. The scaling procedure
allows to extract the stationary part, but not to reach the
limit where a full time-scale separation is achieved.

A way to reach the limit could be to iterate the scaling
procedure on the ageing data to separate the ``ageing timescales''. The ageing regime can be considered
as a pseudo-FDT one, with a temperature equal to
$T_{eff}(q_{EA})$. Such a work on the available data is
however hopeless, as the separation between stationary and
ageing part seems obviously already far from perfect.

A scaling can be deduced from equation~\ref{eq:master_extended} \cite{Marinari00}, which, using
$\Phi={1 \over 1-B}$, can be written as, :

\begin{equation}
  \chi.T^{1-\phi} = \left\{
    \begin{array}{ll}
      A[(1-C)T^{-\Phi}]^B & {\rm for} \;\; C \le q_{EA}(T) \ , \\
      (1-C)T^{-\Phi} & {\rm for} \;\; C > q_{EA}(T) \ .
    \end{array}
  \right.
  \label{eq:ScalingT}
\end{equation}

If a power-law can describe the ageing dynamics, then all
the scaled data should merge along a single line. The best
result is obtained for $B=0.5$, but the cloud of points
remains very broad, and is not well described by the
predicted straight-line in the $\log(T^{1-\Phi}\chi)$ vs
$\log\left(T^{-\Phi}(1-C)\right)$ diagram.

\begin{figure}[htb]
  \centering
  \includegraphics[width=0.45\textwidth]{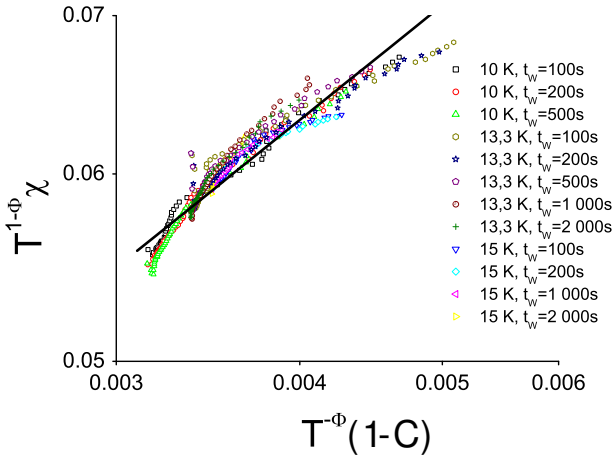}
  \caption{Scaling of the ageing parts of the FDT-diagram following equation
  \ref{eq:ScalingT}. The straight line shows the result predicted by this equation.}
  \label{fig:ScalingT}
\end{figure}

\section{Conclusion}
\label{sec:concl} In this paper, it has been shown that the
experimental setup developed in this work can be considered
as an absolute magnetic thermometer. However, to get rid of
uncertainties on the value of several elements of the
setup, the setup was calibrated by measuring the magnetic
fluctuations and response of a high conductivity copper
sample thermalised at 
helium temperature. This calibrated thermometer was used to
determine the out-of-equilibrium properties of a spin-glass
below the glass transition. The autocorrelation function of
the spontaneous magnetic fluctuations of a well
characterised insulating spin-glass was investigated in the
deeply non-stationary regime. Its waiting time dependence
can be described by using the same scaling as for the
response function. The FD-plots clearly confirm that
the stationary dynamics observed at the shorter timescales
can be considered as a quasi-equilibrium once, as the
fluctuation-dissipation relation between autocorrelation
and relaxation obeys the fluctuation-dissipation theorem. 
The results show clearly that the asymptotic regime, with full separation of timescale is not reached. Certain Hypotheses on the dynamics are made  in order to compare the results with model predictions.
The deduced scaling analysis allows to extrapolate the experimental
results to the limit used in theoretical studies of
weak-ergodicity breaking models. 

The experimental results obtained on \thio{ } differ
qualitatively from the predictions of any domain-growth
model.
 
 The experimental data allow interpretations that are rather consistent with predictions from the two replica symmetry breaking models under study. As long as the autocorrelation
cannot be determined completely, both models
can be relevant, giving slightly different results
concerning the value of $q_{EA}$. An independent determination of the
characteristic magnetic moment of the clusters as a
function of temperature is needed in order to resolve this
ambiguity .

The possibility of analysing the experimental results on
the basis of 1-step replica symmetry breaking confirms
that, at first sight, the chiral model developed by
Kawamura could be the more relevant one for the \thio
compound with low anisotropy, supporting the conclusion from D. Petit and I. Campbell on this sample \cite{Petit00}.
However, the scatter of the data is such that one cannot
reject definitely an interpretation inspired by the
mean-field models developed for Ising spin-glasses.

\section{Acknowledgements}
DH has been supported by the European Research Training Network \textsc{Dyglagemem} during the writting of this article. We thank G. Parisi for fruitfull discussions and Per Nordblad for a critical reading of the manuscript.
\bibliographystyle{unsrt}
\bibliography{../PyBibliodh}

\end{document}